\documentclass[12pt,a4paper]{article} %UKenglish,
\usepackage[utf8]{inputenc}
\usepackage{bm}
\usepackage{color}
\usepackage{amsmath}
\usepackage{amsthm}
\usepackage{amssymb}
\usepackage{graphicx}
\usepackage{verbatim}
\usepackage[breaklinks]{hyperref}

\newtheorem{lemma}{Lemma}
\newtheorem{theorem}{Theorem}

\newtheorem{claim}{Claim}
\theoremstyle{definition}
\newtheorem{definition}{Definition}
\newtheorem{oldtheorem}{Theorem}

\definecolor{aocolour}{rgb}{0.7,0.8,1}
\definecolor{omcolour}{rgb}{1,0.4,0.4}

\newcommand{\set}[2]{\ensuremath{ \{ \, #1 \mid #2 \, \} }}

\renewcommand{\epsilon}{\varepsilon}

\newcommand{\psinotL}{\mathbb{N}^9 \setminus \psi(L)}

\begin{document}

\title{Non-closure under complementation for unambiguous linear grammars} %\thanks{%
\author{Olga Martynova\thanks{%
	Department of Mathematics and Computer Science,
	St.~Petersburg State University, 7/9 Universitetskaya nab., Saint Petersburg 199034, Russia,
	\emph{ and }
	Leonhard Euler International Mathematical Institute at St.~Petersburg State University,
	Saint Petersburg, Russia.
	E-mail: \texttt{olga22mart@gmail.com}
} \and Alexander Okhotin\thanks{%
	Department of Mathematics and Computer Science,
	St.~Petersburg State University, 7/9 Universitetskaya nab., Saint Petersburg 199034, Russia,
	E-mail: \texttt{alexander.okhotin@spbu.ru}
}}
\maketitle

\begin{abstract}
The paper demonstrates the non-closure
of the family of unambiguous linear languages
(that is, those defined by unambiguous linear context-free grammars)
under complementation.
To be precise, a particular unambiguous linear grammar is presented,
and it is proved that the complement of this language
is not defined by any context-free grammar.
This also constitutes an alternative proof
for the result of Hibbard and Ullian
(``The independence of inherent ambiguity from complementedness among context-free languages'', \emph{JACM}, 1966)
on the non-closure of the unambiguous languages under complementation.
%
%Context-free grammars \sep unambiguous grammars \sep linear grammars \sep complementation \sep closure properties
\end{abstract}

\sloppy

\section{Introduction}

Closure properties of the basic families of formal grammars
are among the textbook results in theoretical computer science:
it is well-known that the complement of a context-free language
need not be context-free,
whereas the complement of an LR(1) language is always LR(1), etc.
However, the impression of knowing completely everything is false,
and there are in fact a few unsolved questions of this kind
for quite basic grammar families,
see the list in Table~\ref{t:closure_properties}.
This paper addresses one of these questions---%
whether the family of \emph{unambiguous linear languages}
is closed under complementation---%
and settles it in the negative.

The family of \emph{unambiguous grammars}
is known since Chomsky and Sch\"utzenberger~\cite{ChomskySchutzenberger}:
these are context-free grammars that define a unique parse tree
for every string they generate.
In the subfamily of \emph{unambiguous linear grammars},
the right-hand side of every rule contains at most one nonterminal symbol.
Unambiguous linear grammars have received some attention on their own:
Ceccherini-Silberstein~\cite{Ceccherinisilberstein} presented an algorithm
for determining the growth rate of a language defined by such a grammar,
while Diekert et al.~\cite{DiekertKopeckiMitrana} proved that the hairpin completion of a regular language
is always defined by a grammar from this class.

The language family defined by the whole class of unambiguous grammars
is known not to be closed under most standard operations.
Proving that requires showing
non-representability of particular languages by these grammars,
and several proof methods have been developed for this purpose.

\newcommand{\family}[1]{\mbox{\textsf{\small \textup{\textit{#1}}}}}
\newlength{\widthofthisref}
\newcommand{\mycef}[2]{\settowidth{\widthofthisref}{\footnotesize #2}%
	\hspace*{\widthofthisref}\:#1\:\text{\footnotesize #2}}
\newcommand{\myref}[2]{#1}
\newcommand{\mylef}[2]{#1}
\begin{table*}[t]
	\centerline{\(
\begin{array}{|l|cccccc|}
	\hline
	&\cup	
		&\cap	
			&\sim
				&\cdot
					&*
						&R
									\\
	\hline
\text{Regular}
	&+
		&+
			&+
				&+
					&+
						&+
									\\
\text{Linear LL($k$)}
	&\myref{-}{\cite{RosenkrantzStearns}} % Thm 10: a^n b^n \cup a^n c^n
		&\myref{-}{\cite{BooleanLLfamily}} % E1
			&\myref{-}{\cite{RosenkrantzStearns}} % Cor 4
				&\myref{-}{\cite{RosenkrantzStearns}} % Thm 10 a^* \cdot a^n b^n
					&\myref{-}{\cite{Wood_further}} % L55
						&\myref{-}{\cite{RosenkrantzStearns}} % Thm 10 (b^n a^n \cup c^n a^n)^R
									\\
\text{Linear LR(1)}
	&-
		&- % a^m c^* b^m \cap a^* b^n c^n \notin CF
			&+
				&- %\{a^n b^n\}^2
					&- %\{a^n b^n\}^*
						&\myref{-}{\cite{GinsburgGreibach}} %p.641
									\\
\text{Unambiguous linear}
	&- % a^m c^* b^m \cup a^* b^n c^n \notin UnambCF
		&- % a^m c^* b^m \cap a^* b^n c^n \notin CF
			&\text{\textbf{?}} %\myref{-}{\cite{UnamblinComplement}}
				&- %\{a^n b^n\}^2
					&- %\{a^n b^n\}^*
						&+
									\\
\text{Linear}
	&+
		&-
			&-
				&-
					&-
						&+
								\\
\text{LL($k$)}
	&\myref{-}{\cite{RosenkrantzStearns}} % Thm 10
		&\myref{-}{\cite{RosenkrantzStearns}} % Thm 10
			&\myref{-}{\cite{RosenkrantzStearns}} % Thm 10
				&\myref{-}{\cite{RosenkrantzStearns}} % Thm 10
					&\myref{-}{\cite{Wood_further}}
						&\myref{-}{\cite{RosenkrantzStearns}} % Thm 10
									\\
\text{LR(1)}
	&\myref{-}{\cite{GinsburgGreibach}}
		&\myref{-}{\cite{GinsburgGreibach}}
			&\myref{+}{\cite{GinsburgGreibach}}
				&\myref{-}{\cite{GinsburgGreibach}}
					&\myref{-}{\cite{GinsburgGreibach}}
						&\myref{-}{\cite{GinsburgGreibach}} %p.641
									\\
\text{Unambiguous}
	&-
		&-
			&\myref{-}{\cite{HibbardUllian}} %UnamblinComplement}} % They have a really difficult subset of a^* b^* c^* d^*, mine is better, but the proof is not elementary
				&\myref{-}{\cite{GinsburgUllian_preservation}} % Thm 4, with a 2-element set.
					&- %from non-closure under concatenation, via (cL_1 \cup L_2c)^* \cap c\Sigma^*c
						&+
								\\
\text{Ordinary (context-free)}
	&+
		&\myref{-}{\cite{Scheinberg}}
			&\myref{-}{\cite{Scheinberg}}
				&+
					&+
						&+
								\\
	\hline
\end{array}
\)}
	\caption{Closure properties of different families of formal grammars.}
	\label{t:closure_properties}
\end{table*}

First, there are combinatorial methods based on Ogden's lemma,
and they are sufficient to prove that the unambiguous languages
are not closed under union and concatenation~\cite{GinsburgUllian_preservation}.

Another powerful proof method is based on the fact
that the generating function of every unambiguous language is algebraic~\cite{ChomskySchutzenberger}.
Flajolet~\cite{Flajolet}, in his famous paper,
used this method to prove inherent ambiguity of numerous languages.
However, this method cannot be used to prove non-closure under complementation,
because if the generating function of a language is algebraic,
then so is the generating function of its complement.

A recent algebraic method for proving non-existence of an unambiguous grammar for a given language,
introduced by Makarov~\cite{Makarov},
is based on showing a stronger result
that there is no \emph{GF(2)-grammar} for that language.
But this whole method is again of no use for showing non-closure under complementation,
because this operation is effectively representable in GF(2)-grammars.

The non-closure of the unambiguous languages under complementation
was proved by Hibbard and Ullian~\cite{HibbardUllian},
who constructed a ``bounded'' language $L \subseteq a^* b^* c^* d^*$,
and used the method of Ginsburg and Spanier~\cite{GinsburgSpanier_bounded,GinsburgSpanier_presburger}
to show that no context-free grammar defines its complement.
The argument involved representing strings in $a^* b^* c^* d^*$
as points in a four-dimensional vector space,
and using an essentially geometrical argument to show that the complement of a certain set
cannot be covered by effectively representable subsets.

This paper establishes a stronger result:
the new witness language is defined by an \emph{unambiguous linear grammar},
yet its complement is not defined by any context-free grammar.
This proves the non-closure of the unambiguous linear languages
under complementation,
as well as provides an alternative proof
for the result of Hibbard and Ullian~\cite{HibbardUllian}.

\section{Grammars and linear sets}

This paper uses the ordinary (``context-free'') formal grammars,
referred to simply as \emph{grammars},
as well as their linear and unambiguous subfamilies,
defined in the usual way in terms of parse trees.

\begin{definition}
A grammar is a quadruple $G=(\Sigma, N, R, S)$,
where $\Sigma$ is the alphabet of the language being defined,
$N$ is the set of nonterminal symbols,
$R$ is a finite set of rules, each of the form $A \to \alpha$,
with $A \in N$ and $\alpha \in (\Sigma \cup N)^*$,
and $S \in N$ is the initial symbol.

A parse tree is a finite tree,
with each node labelled with a symbol from $\Sigma \cup N$,
and with the successors of each node linearly ordered.
For each node labelled with a nonterminal symbol $A \in N$,
the labels of its direct successors
must form the right-hand side of one of the rules for $A$.
The root node is labelled with $S$.
A node labelled with a symbol from $\Sigma$ has no successors.
The yield of a tree is a string $w \in \Sigma^*$
formed by the labels of the leaves, listed in order;
then, the tree is called a parse tree of $w$.
The language defined by the grammar, denoted by $L(G)$,
is the set of all strings $w$
which have at least one parse tree.
\end{definition}

\begin{definition}
A grammar $G=(\Sigma, N, R, S)$ is called unambiguous
if every string has at most one parse tree.
\end{definition}

\begin{definition}
A grammar $G=(\Sigma, N, R, S)$ is called linear,
if the right-hand side of every rule in $R$
contains at most one nonterminal symbol,
that is, all rules are of the form
$A \to uBv$, with $u, v \in \Sigma^*$ and $B \in N$,
or $A \to w$, with $w \in \Sigma^*$.
\end{definition}

A language is \emph{unambiguous linear}
if it is defined by a grammar that is both unambiguous and linear.
The problem addressed in this paper is whether the complement of every unambiguous linear language
is always unambiguous linear,
and a negative answer is given.

A witness language, which is unambiguous linear, whereas its complement is not,
is going to be a subset of $a_1^* \ldots a_k^*$,
defined over an alphabet $\Sigma=\{a_1, \ldots, a_k\}$;
this is the basic case of \emph{bounded languages}
of Ginsburg and Spanier~\cite{GinsburgSpanier_bounded}.
Strings of this form are in one-to-one correspondence with $\mathbb{N}^k$,
where $\mathbb{N}=\{0, 1, 2, \ldots\}$ is the set of non-negative integers.

Denote this correspondence by
$\psi(a_1^{i_1}, \ldots, a_k^{i_k})=(i_1, \ldots, i_k) \in \mathbb{N}^k$.
The image of a language $L \subseteq a_1^* \ldots a_k^*$
is the set
$\psi(L)=\set{\psi(w)}{w \in L} \subseteq \mathbb{N}^k$.
These languages have a convenient representation in terms of some basic linear algebra.

\begin{definition}%[Linear sets]
A set $S \subseteq \mathbb{N}^k$ is called \emph{linear}
if it is representable as $\set{\alpha + \sum_{i=1}^m x_i \beta_i}{x_i \in \mathbb{N}}$,
for some basis vectors $\beta_1, \ldots, \beta_m \in \mathbb{N}^k$
and a shift vector $\alpha \in \mathbb{N}^k$.
\end{definition}

Ginsburg and Spanier~\cite{GinsburgSpanier_bounded}
singled out a subclass of linear sets representable by grammars,
which are subject to the following condition.

\begin{definition}\label{definition_stratified_linear_set}
A linear set
$\set{\alpha + \sum_{i=1}^m x_i \beta_i}{x_i \in \mathbb{N}} \subseteq \mathbb{N}^k$
with $\alpha=(\alpha_1, \ldots, \alpha_k)$
and $\beta_i=(\beta_{i,1}, \ldots, \beta_{i,k})$
is said to be \emph{stratified} if
every basis vector $\beta_i$ has at most two nonzero coordinates,
and there do not exist $\beta_i$ and $\beta_{i'}$
and coordinates $1 \leqslant j_1 < j_2 < j_3 < j_4 \leqslant k$,
such that $\beta_{i,j_1},\beta_{i',j_2},\beta_{i,j_3},\beta_{i',j_4} \neq 0$.
\end{definition}

\begin{oldtheorem}[Ginsburg and Spanier {\cite[Thm.2.1]{GinsburgSpanier_presburger}}]\label{bounded_language_stratified_linear_theorem}
A language $L \subseteq a_1^* \ldots a_k^*$
over $\Sigma=\{a_1, \ldots, a_k\}$
is defined by a grammar
	if and only if
$\psi(L)$ is a finite union of stratified linear sets.
\end{oldtheorem}

\section{The non-closure result}

This section presents the result of this paper:
a language $L$ witnessing the non-closure of the unambiguous linear languages
under complementation.

The language $L$ is defined
over a 9-symbol alphabet $\Sigma = \{a_1, a_2, \ldots, a_9\}$.
It is a bounded language, with $L \subseteq a_1^* \ldots a_9^*$,
defined as a union of the following four languages.
\begin{align*}
	L_1 &= \set{a_1^{i_1} \ldots a_9^{i_9}}{i_1 \leqslant i_9, \; i_2 \leqslant i_7, \; i_3 \leqslant i_5}, \\
	L_2 &= \set{a_1^{i_1} \ldots a_9^{i_9}}{i_1 > i_9, \; i_2 \leqslant i_6, \; i_3 \leqslant i_4}, \\
	L_3 &= \set{a_1^{i_1} \ldots a_9^{i_9}}{i_1 \leqslant i_8, \; i_2 > i_7, \; i_3 > i_4}, \\
	L_4 &= \set{a_1^{i_1} \ldots a_9^{i_9}}{i_1 > i_8, \; i_2 > i_6, \; i_3 > i_5}.
\end{align*}
Each language is defined by three comparisons,
which always involve the first three coordinates:
the first coordinate $i_1$ is compared with either $i_8$ or $i_9$,
the second coordinate $i_2$ is compared with either $i_6$ or $i_7$,
and the third coordinate $i_3$ is compared with either $i_4$ or $i_5$.
For each language, these three comparisons are always well-nested,
and hence can be defined by a linear grammar (which is also unambiguous, and is presented below).
The comparison between two particular coordinates, such as $i_1$ and $i_9$,
is made in exactly two different languages:
in one of them, $i_1 \leqslant i_9$, and $i_1 > i_9$ in the other.
Every two languages are thus separated by exactly one common pair of coordinates:
for example, $L_1$ and $L_2$ are separated
by whether $i_1$ is less or greater than $i_9$.
Therefore, these four languages are pairwise disjoint.

Since each of these four languages
is defined by an unambiguous linear grammar,
their union is defined by such a grammar as well.
However, the complement of their union
is not defined by any grammar at all.

\begin{theorem}\label{unamblin_complement_theorem}
The language $L = L_1 \cup L_2 \cup L_3 \cup L_4$ is defined by an unambiguous linear grammar,
whereas its complement is not defined by any (context-free) grammar.
\end{theorem}

Constructing an unambiguous linear grammar
for each of the languages $L_1$, $L_2$, $L_3$ and $L_4$
is an exercise.
For instance, here is the grammar for
$L_1 = \set{a_1^{i_1} \ldots a_9^{i_9}}{i_1 \leqslant i_9, \; i_2 \leqslant i_7, \; i_3 \leqslant i_5}$.
\begin{align*}
	S_1 &\to A_{1,9} \\
	A_{1,9} &\to a_1A_{1,9}a_9 \ | \ A_9 \\
	A_9 &\to A_9a_9 \ | \ A_8 \\
	A_8 &\to A_8a_8 \ | \ A_{2,7} \\
	A_{2,7} &\to a_2A_{2,7}a_7 \ | \ A_7 \\
	A_7 &\to A_7a_7 \ | \ A_6 \\
	A_6 &\to A_6a_6 \ | \ A_{3,5} \\
	A_{3,5} &\to a_3A_{3,5}a_5 \ | \ A_5 \\
	A_5 &\to A_5a_5 \ | \ A_4 \\
	A_4 &\to A_4a_4 \ | \ \epsilon
\end{align*}
%Аналогичным образом задаются $L_2$, $L_3$, $L_4$.
The other three grammars are constructed similarly.
The union of these four languages is defined by a grammar
obtained by combining the grammars for $L_1$, $L_2$, $L_3$ and $L_4$,
and adding a new initial symbol $S$
and new rules $S \to S_1 \ | \ S_2 \ | \ S_3 \ | \ S_4$.
Since these four languages are disjoint, the resulting grammar remains unambiguous.

To see that there is no grammar for the complement of $L$,
suppose, for the sake of a contradiction, that $\Sigma^*\setminus L$ is defined by some grammar.
Then, the following intersection with a regular language is also defined by some grammar.
\begin{equation*}
	(\Sigma^*\setminus L)
	\cap
	a_1^* a_2^* \ldots a_9^*
\end{equation*}
By Theorem~\ref{bounded_language_stratified_linear_theorem},
the set $\psi\big((\Sigma^*\setminus L) \cap a_1^* a_2^* \ldots a_9^*\big)$,
which equals $\mathbb{N}^9 \setminus \psi(L)$,
is a finite union of stratified linear sets.
In order to prove Theorem~\ref{unamblin_complement_theorem},
it is left to show that $\mathbb{N}^9 \setminus \psi(L)$
is not representable as such a union.
In fact, a stronger result will be obtained:
that $\mathbb{N}^9 \setminus \psi(L)$ is not a finite union of any linear sets
with at most two non-zero coordinates in each basis vector
(whether stratified or not).

\begin{lemma}\label{psi_not_L_not_finite_union_lemma}
The set $\psinotL$
is not a finite union of linear sets of the form
$\set{\alpha + \sum_{i=1}^m x_i \beta_i}{x_i \in \mathbb{N}} \subseteq \mathbb{N}^9$,
in which every basis vector $\beta_i$ has at most two non-zero coordinates.
\end{lemma}

Note that this is a stronger condition than
non-representability by a finite union of stratified linear sets,
because the non-crossing condition in Definition~\ref{definition_stratified_linear_set}
is not used.

\begin{proof}
In this proof, linear sets
with basis vectors restricted to have at most two non-zero coordinates each
are referred to as \emph{light linear sets}.
Suppose $\psinotL$ is represented as a union
of finitely many light linear sets,
and let $M$ be the smallest integer that is strictly greater
than all coordinates of all basis vectors and shift vectors in these sets.

Consider the following point $v$ in the nine-dimensional space.
\begin{equation*}
\begin{array}{cccccccccc}
	& v_1 & v_2 & v_3 & v_4 & v_5 & v_6 & v_7 & v_8 & v_9 \\
v = 	& (M, & 3M, & 2M, & 2M, & M, & 2M, & 2M, & M, & M)
\end{array}
\end{equation*}

The point $v$ is not in $\psi(L)$:
indeed, it is not in $\psi(L_1)$, because $v_2 > v_7$;
not in $\psi(L_2)$, because $v_2 > v_6$;
not in $\psi(L_3)$, because $v_3 = v_4$;
not in $\psi(L_4)$, because $v_1 = v_8$;

Since $v \notin \psi(L)$,
it is covered by some light linear set $S_v \subseteq \psinotL$.
Let $S_v = \set{\alpha + \sum\limits_{i = 1}^m{c_i\beta^i}}{c_1, \ldots, c_m \in \mathbb{N}}$, 
where $\alpha = (\alpha_1,\ldots, \alpha_9)$ is the shift vector,
$\beta^i =  (\beta^i_1,\ldots, \beta^i_9)$, with $1 \leqslant i \leqslant m$, are basis vectors,
and each $\beta^i$ has at most two non-zero coordinates.
Some linear combination of the vectors $\beta^1,\ldots, \beta^m$,
shifted by the vector $\alpha$,
gives the point $v = (v_1, \ldots, v_9)$.
If, in this representation of $v$,
some of the basis vectors are included with zero coefficients,
then the corresponding vectors can be removed from $S_v$,
and $S_v$ remains a light linear subset of $\psinotL$,
which contains $v$.
For this reason, assume that in the representation $v=\alpha + \sum_{i=1}^m c_i \beta^i$,
all coefficients $c_1, \ldots, c_m$ are positive.

No points of the set $S_v$ are in $\psi(L)$,
since $S_v \subseteq \psinotL$.
The contradiction needed to prove the lemma
shall be obtained by moving the point $v$,
so that it remains in the set $S_v$,
but, at the same time, ends up in $\psi(L)$.

\begin{claim}\label{claim_1_8}
Every basis vector $\beta^i$, with $1 \leqslant i \leqslant m$,
has the same 1st and 8th coordinates: $\beta^i_1=\beta^i_8$.
Furthermore, there exists at least one basis vector $\beta^{1,8} \in \{\beta^1, \ldots, \beta^m\}$
with a non-zero pair in these coordinates: $\beta^{1,8}_1=\beta^{1,8}_8 > 0$.
\end{claim}
\begin{proof}
The point $v$ is not in $\psi(L_4) = \set{(i_1, \ldots, i_9)}{i_1 > i_8, \; i_2 > i_6, \; i_3 > i_5}$ 
only because $v_1=v_8 = M$;
the other two conditions of membership do hold ($3M = v_2 > v_6 = 2M$, $2M = v_3 > v_5 = M$).
If there exists at least one vector $\beta^i$ with different 1st and 8th coordinates,
then the point $v$ can be driven into $\psi(L_4)$
by increasing or decreasing the coefficient $c_i$ at $\beta^i$ by 1,
so that the 1st coordinate becomes greater than the 8th coordinate;
at the same time,
the addition or subtraction of a single vector
would not violate the inequalities $v_2 > v_6$ and $v_3 > v_5$,
because the values of all coordinates of $\beta^i$ do not exceed $M-1$,
whereas the differences $v_2 - v_6$ and $v_3 - v_5$ are both equal to $M$.

To see that there exists a basis vector $\beta^{1,8}$
with non-zero 1st and 8th coordinates.
consider that the contribution of the shift vector $\alpha$ to each coordinate is less than $M$,
whereas $v_1 = v_8 = M$.
Then, there must exist a basis vector with non-zero 1st or 8th coordinate.
Since these two coordinates are equal in all basis vectors,
this vector is the desired vector $\beta^{1,8}$.
The rest of the coordinates of $\beta^{1,8}$ are zero,
since every basis vector has at most two non-zero coordinates.
\end{proof}

\begin{claim}\label{claim_3_4}
Every basis vector $\beta^i$, with $1 \leqslant i \leqslant m$,
has the same 3rd and 4th coordinates: $\beta^i_3=\beta^i_4$.
\end{claim}
\begin{proof}
The point $v$ is not in $\psi(L_3) = \set{(i_1, \ldots, i_9)}{i_1 \leqslant i_8, \; i_2 > i_7, \; i_3 > i_4}$
only because $v_3 = v_4 = 2M$;
the other two conditions hold true: $v_1 = v_8 = M$ and $3M = v_2 > v_7 = 2M$.
If some basis vector $\beta^i$ has different 3rd and 4th coordinates,
then, by increasing or decreasing its coefficient $c_i$ by 1,
one can make the 3rd coordinate greater than the 4th one.
This adjustment of $c_i$ preserves the equality $v_1 = v_8$,
because, by Claim~\ref{claim_1_8}, the vector $\beta^i$
has equal 1st and 8th coordinates (as all basis vectors have).
The inequality $v_2 > v_7$ also still holds, since $v_2-v_7 = M$,
while all coordinates of $\beta^i$ are less than $M$.
This confirms that if any basis vector $\beta^i$ has $\beta^i_3 \neq \beta^i_4$,
then $S_v$ contains a point from $\psi(L_3)$, contrary to the definition of $S_v$. 
\end{proof}

Now consider a point $u$,
which is obtained from $v$
by removing all basis vectors with non-zero 3rd and 4th coordinates
from the shifted linear combination $v = \alpha+\sum_{i = 1}^mc_i\beta^i$
(that is, changing the coefficients of all such vectors to zeroes).
Every vector thus removed has all coordinates except the 3rd and the 4th equal to zero,
because a basis vector has at most two non-zero coordinates.
Accordingly, $u$ differs from $v$ only in the 3rd and the 4th coordinates.
After this change, the 3rd and the 4th coordinates are still equal ($u_3=u_4$),
because they were equal in the original point ($v_3 = v_4$),
and every vector removed has equal 3rd and 4th coordinates by Claim~\ref{claim_3_4}.
In all basis vectors remaining in $u$, the 3rd and the 4th coordinates are zeroes,
and hence only the shift vector $\alpha$ contributes to $u_3$ and $u_4$.
Therefore, $u_3 = u_4 < M$,
and the coordinates of the point $u$ are as follows.

\begin{equation*}
\begin{array}{cccccccccc}
	& u_1 & u_2 & u_3 & u_4 & u_5 & u_6 & u_7 & u_8 & u_9 \\
u = 	& (M, & 3M, & x<M, & x<M, & M, & 2M, & 2M, & M, & M)
\end{array}
\end{equation*}

\begin{claim}\label{claim_2_6}
Basis vectors with the 2nd and the 6th coordinates both non-zero
contribute less than $M$ to $u_2$.
\end{claim}
\begin{proof}
The point $u$ is not in $\psi(L_1) = \set{(i_1, \ldots, i_9)}{i_1 \leqslant i_9, \; i_2 \leqslant i_7, \; i_3 \leqslant i_5}$
only because $u_2 > u_7$;
two other conditions do hold, since $u_1=u_9=M$ and $x=u_3 < u_5 = M$.
If basis vectors with non-zero 2nd and 6th coordinates
contribute at least $M$ to $u_2$,
then let us remove those vectors from the shifted linear combination that defines $u$.
This affects only $u_2$ and $u_6$,
and $u_2$ is reduced at least by $M$,
so that it becomes at most $2M$, and accordingly, not greater than $u_7$.
The other two conditions of membership in $\psi(L_1)$ are unaffected,
because only the coordinates $u_2$ and $u_6$ were modified.
Then the resulting point is in $\psi(L_1)$, which cannot be the case.
The contradiction obtained proves the claim.
\end{proof}

Now consider another point $w \in S_v$,
which is obtained from $u$ in two steps.
First, all basis vectors with a positive 2nd coordinate and at the same time zero 6th coordinate
are removed from the shifted linear combination defining $u$.
Secondly, the coefficient at the basis vector $\beta^{1,8}$---%
a vector with positive 1st and 8th coordinates that exists by Claim~\ref{claim_1_8}---%
is increased by $M+1$.

It is claimed that $w \in \psi(L_2)$, which will yield a contradiction.
All three inequalities from the definition 
of $\psi(L_2) = \set{(i_1, \ldots, i_9)}{i_1 > i_9, \; i_2 \leqslant i_6, \; i_3 \leqslant i_4}$
need to be checked.
\begin{itemize}
\item
	The inequality $w_1 > w_9$ holds,
	because removing some basis vectors may only decrease the 9th coordinate,
	and $w_9 \leqslant M$,
	whereas $w_1 > M$, as the coefficient at $\beta^{1,8}$ 
	has been increased by $M+1$.
\item
	The inequality $w_2 \leqslant w_6$ is verified as follows.
	Since only basis vectors with a zero 6th coordinate were removed,
	while the vector $\beta^{1,8}$ has only the 1st and the 8th coordinates non-zero,
	the 6th coordinate is unaffected: $w_6 = u_6 = 2M$.
	The coordinate $w_2$ is made of contributions by the shift vector,
	and by basis vectors with non-zero 2nd coordinate.
	The 6th coordinate of these basis vectors are non-zero,
	because those with a zero 6th coordinate were removed by the construction of $w$.
	By Claim~\ref{claim_2_6}, the contribution of vectors with non-zero 2nd and 6th coordinates
	to $u_2$, and therefore to $w_2$, does not exceed $M$. 
	The contribution of the shift vector to $w_2$ is less than $M$.
	Therefore, $w_2 < 2M = w_6$.
\item
	Finally, $w_3 = w_4$, because $u_3 = u_4$, 
	and $w$ differs from $u$ only in the coefficients at some basis vectors,
	while the 3rd and the 4th coordinates of each basis vector are equal by Claim~\ref{claim_3_4}.
\end{itemize}

Thus, the point $w$, which is in $S_v \subseteq \psinotL$,
satisfies all conditions in the definition of $\psi(L_2) \subseteq \psi(L)$.
This is a contradiction.
\end{proof}

\section{Conclusion}

This settles one of the last few remaining closure properties
of basic families of formal grammars.

Turning to extensions of the context-free grammars,
there are still some open problems on their closure properties.
For instance, for grammars equipped with a conjunction operation---%
\emph{conjunctive grammars}~\cite{Conjunctive,ConjunctiveTokyo}---%
it is not known whether their language family is closed under complementation,
and the methods based on linear sets are unlikely to help settling this problem,
because conjunctive grammars can express bounded languages
beyond linear sets and unions thereof.

Another interesting question refers to \emph{multi-component grammars},
also known as \emph{multiple context-free grammars}~\cite{SekiMatsumuraFujiiKasami}.
Their non-closure under complementation is known~\cite[Sect.~3.1]{SekiMatsumuraFujiiKasami},
but the same problem for their unambiguous subclass is apparently open.
Perhaps a non-closure result for this family could be established
by adapting the methods involving linear sets.

\section*{Acknowledgement}

This work was supported by
the Ministry of Science and Higher Education of the Russian Federation,
agreement 075-15-2019-1619.

\end{document}